\documentclass[12pt,amsbsy]{article}
\usepackage[centertags]{amsmath}
\usepackage[cp866]{inputenc}
\usepackage{graphicx}
\newcommand{\beq}{\begin{equation}\label{}}
\newcommand{\eeq}{\end{equation}}
\newcommand{\beqn}{\begin{eqnarray}}
\newcommand{\eeqn}{\end{eqnarray}}

\newcommand{\ga}{\mbox{${\gamma}$}}
\newcommand{\Ga}{\mbox{${\Gamma}$}}
\newcommand{\de}{\mbox{${\delta}$}}
\newcommand{\De}{\mbox{${\Delta}$}}
\newcommand{\ka}{\mbox{${\kappa}$}}

\newcommand{\om}{\mbox{${\omega}$}}

\begin{document}

\begin{center}
{\bf \Large Quasinormal modes for arbitrary spins\\ in the
Schwarzschild background}

\vspace{5mm}

I.B. Khriplovich\footnote{khriplovich@inp.nsk.su} and G.Yu.
Ruban\footnote{gennady-ru@ngs.ru}

\vspace{3mm}

Budker Institute of Nuclear Physics\\ 630090 Novosibirsk, Russia,\\
and Novosibirsk University
\end{center}

\vspace{5mm}

\begin{abstract}

The leading term of the asymptotic of quasinormal modes in the
Schwarz\-schild background, $\om_n = - \,i\, n/2$, is obtained in
two straightforward analytical ways for arbitrary spins. One of
these approaches requires almost no calculations. As simply we
demonstrate that for any odd integer spin, described by the
Teukolsky equation, the first correction to the leading term
vanishes. Then, this correction for half-integer spins is
obtained in a slightly more intricate way. At last, we derive
analytically the general expression for the first correction for
all spins, described by the Teukolsky equation.

\end{abstract}

\section{Introduction}
The investigation of perturbations of various fields in the
Schwarzschild background was started in \cite{RW,ze}. Quasinormal
modes (QNM) are the eigenmodes of the homogeneous wave equations,
describing these perturbations, with the boundary conditions
corresponding to outgoing waves at the spatial infinity and
incoming waves at the horizon. The interest to QNMs was initiated
by \cite{vi,pre}.

Two boundary conditions make the frequency spectrum $\om_n$ of
QNMs discrete. The asymp\-totic form of this spec\-trum for
gravi\-tational and scalar pertur\-bations of the Schwarz\-schild
background was found at first numerically in \cite{lea,nol}:
\beq\label{o02}
\om_n = \, -\, \frac{i}{2}\, \left(n+\,\frac{1}{2}\right) +\,
0.087424\,, \quad n \to \infty\,, \quad s\,=\,0,\,2\,.
\eeq
Here and below the gravitational radius $r_g$ is put to unity; $s$
is the spin of the perturbation. This result up to now serves as a
touch stone for investigations in the field.

A curious observation was made in \cite{hod}: the real constant in
(\ref{o02}) can be presented as
\beq\label{rom}
\rm{Re}\, \om_{\it n} = \,\frac{\ln 3}{4\pi}\,= {\it T_H}\, \ln 3,
\eeq
where $T_H$ is the Hawking temperature ($T_H=1/(8\pi k M)$ in the
common units)\footnote{It was also conjectured in \cite{hod} that
the asymptotic value (\ref{rom}) for $\,\rm{Re}\,\om_{\it n}$ is
of a crucial importance for the quantization of gravitational
field, fixing the value of the so-called Barbero -- Immirzi
parameter. In spite of being very popular, this idea is not in
fact dictated by any sound physical arguments; quite the contrary,
it is in conflict with them \cite{KHI1}.}. Then, expression
(\ref{rom}) for the asymptotic of $\rm{Re}\, \om_{\it n}$ was
derived in~\cite{mot} by solving approximately the recursion
relations used previously in the numerical calculations. In the
next paper~\cite{mon} formula (\ref{rom}) was derived
analytically. Besides, in \cite{mot} the following result was
obtained for the asymptotic of QNMs for spin~$1$:
\beq\label{o1}
\om_{\it n} = \, -\, \frac{i}{2}\, n\,, \quad \rm{Re}\, \om_{\it
n} \to 0\,, \quad {\it n} \to \infty\,, \quad {\it s}\,=\,1\,.
\eeq
Asymptotic (\ref{o1}) was also obtained numerically in
\cite{car}.

While the results (\ref{o02}) and (\ref{o1}) for integer $s$ are
firmly established now, it is not the case for spin $1/2$. Two
different approaches\footnote{We believe that one of them, despite
being rather popular, can be dismissed at once. It is based on the
analysis of the location of the poles of the scattering amplitude,
which by itself causes no objections. However, following
\cite{pad}--\cite{roy}, the authors of \cite{kon} analyze the
poles of the corresponding Born amplitude. Meanwhile, the Born
approximation by itself implies that the amplitude of the
scattered wave is small. Therefore, its poles have no real
meaning. Any coincidence between their position and that of the
poles of a true amplitude is an accident only.} used in~\cite{kon}
result in the interval two times smaller than those for integer
spins, namely:
\beq\label{kon}
\om_{\it n} = \, -\, \frac{i}{4}\,n\,, \quad n \to \infty\,, \quad
s\,=\,1/2\,.
\eeq
On the other hand, numerical calculations in \cite{jin} result in
spectrum
\beq\label{o12}
\om_{\it n} = \, -\, \frac{i}{2}\,n\,, \quad \rm{Re}\, \om_{\it n}
\to 0\,, \quad {\it n} \to \infty\,, \quad {\it s}\,=\,1/2\,.
\eeq
One of the motivations of our work was the resolution of this
discrepancy; we not only confirm below equation~(\ref{o12}), but
find also first nonvanishing correction to it.

We consider the QNM problem in various analytical approaches. Two
of them, rather simple and straightforward, give in fact only the
leading asymptotic, $\om_{\it n} = -\, i\, n/2$ for any spin. In
the third approach, based on the Teukolsky equation, we at
first demonstrate as easily that equation (\ref{o1}) is accurate
for arbitrary odd integer spins, i.e. that first subleading
correction to it vanishes. Then, with somewhat more efforts, we
obtain this correction for half-integer spins. At last, we derive,
in a more involved way, the unified general expression for the
next term in the asymptotic values of the QNMs for all spins,
which was conjectured previously in \cite{mon}.

\section{Quasinormal modes in Regge -- Wheeler formalism}

The Regge -- Wheeler equation for the radial function $\Psi$
corresponding to the angular momentum $j$ of a field with integer
spin $s$ ($s = 0, 1, 2\,;\;j \geq s$) is written usually as
\begin{equation}\label{rw0}
\frac{d^2\Psi}{dz^2}+\left\{\omega^2-\left(1 -
\frac{1}{r}\right)\left[\frac{j(j+1)}{r^2}+\frac{1-s^2}{r^3}\right]\right\}\Psi=0\,.
\end{equation}
Its analogue for $s=1/2$ (again the angular momentum $j \geq s$),
written for the standard representation of the Dirac
$\ga$-matrices and states of definite parity, is
\begin{equation}\label{rw012}
\frac{d^2\Psi}{dz^2}+\left\{\omega^2- \left(1 -
\frac{1}{r}\right)\frac{(j+1/2)^2}{r^2}\,+\,\frac{\kappa}{2r^3}\left(1
- \frac{1}{r}\right)^{1/2}-\,\frac{\kappa}{r^2}\left(1 -
\frac{1}{r}\right)^{3/2}\right\}\Psi=0\,;
\end{equation}
here $\ka=\pm(j+1/2)$, with the sign depending on the parity of
the state considered (this sign is irrelevant for our problem).
The presence of the terms with fractional powers of $r$ and $r-1$
in equation (\ref{rw012}) is quite natural since wave equations
for half-integer spins are written via tetrads which are roughly
square roots of metric.\footnote{We mention here another rather
popular, but false belief, namely, that equation (\ref{rw0})
applies to half-integer $s$ as well. The explicit difference
between (\ref{rw0}) and (\ref{rw012}) demonstrates that this idea
is wrong.}

In both equations, (\ref{rw0}) and (\ref{rw012}), $r$ is treated
as a function of the so-called ``tortoise'' coordinate $z$. They
are related as follows: $z = r + \ln (r-1)$, so that $z \to
\infty$ for $r \to \infty$, and $z \to -\infty$ for $r \to 1$. The
boundary conditions for QNMs of (\ref{rw0}) and (\ref{rw012}) are
\begin{equation}\label{bc}
\Psi(z) \sim  e^{\pm i\omega z},\quad z\rightarrow \pm\infty.
\end{equation}

Here, for our purpose, it is convenient to go over in both
equations, (\ref{rw0}) and (\ref{rw012}), to the usual coordinate
$r$ and to the new radial function $u(r)$ related to $\Psi$ as
follows:
\beq\label{tr} \Psi = \,\frac{r^{1/2}}{(r-1)^{1/2}}\,u(r)\,. \eeq
The obtained equations for $u(r)$ can be rewritten as
\[
\frac{d^2 u}{dr^2}+\left\{\om^2 + \,\frac{1}{r-1}\left[2\om^2 -
\left(j+\frac{1}{2}\right)^2 +s^2 - \,\frac{1}{4}\right] +
\frac{1}{(r-1)^2}\left(\om^2 + \,\frac{1}{4}\right)\right.
\]
\begin{equation}\label{012}
\quad \left.+\,\frac{1}{r}\left[\left(j+\frac{1}{2}\right)^2 - s^2
+ \frac{1}{4}\right] + \frac{1}{r^2}\left(-s^2 +
\frac{1}{4}\right)\right\}u=0\,, \quad s=0,1,2\,;
\end{equation}
\[
\frac{d^2 u}{dr^2}+\left\{\om^2 + \,\frac{1}{r-1}\left[2\om^2 -
\left(j+\frac{1}{2}\right)^2 +\,\frac{1}{2}\right] +
\frac{1}{(r-1)^2}\left(\om^2 + \,\frac{1}{4}\right)\right.
\]
\[
\left.+\,\frac{1}{r}\left[\left(j+\frac{1}{2}\right)^2 -
\frac{1}{2}\right] -\,\frac{3}{4}\,\frac{1}{r^2}\,
\right.\quad\quad\quad\quad\quad\quad\quad\quad\quad\quad\quad\quad
\]
\begin{equation}\label{12}
\left. -\, \frac{\kappa}{r^{3/2}(r-1)^{1/2}}\,+
\,\frac{1}{2}\,\frac{\kappa}{r^{3/2}(r-1)^{3/2}}\right\}u=0\,,
\quad s=1/2\,.
\end{equation}

We are interested in the solutions of equations (\ref{012}) and
(\ref{12}) in the interval $1~<r~<~\infty$  for $|\,\om| \to
\infty$. Obviously, all the terms singular at $r \to 0$, in both
these equations, are relatively small in this interval if $|\,\om|
\to \infty$.\footnote{In particular, in equation (\ref{12})
\[
|\,\kappa\,|\,r^{-3/2}(r-1)^{-1/2}\, \ll \,|\,\om^2|\, (r-1)^{-1},
\quad {\rm and} \quad |\,\kappa\,|\,r^{-3/2}(r-1)^{-3/2} \ll
|\,\om^2|\,(r-1)^{-2}\,,
\]
for the interval $\,1<r<\infty\,$.}  Therefore, these terms can be
safely omitted, and we arrive at the following universal truncated
wave equation for all spins:
\begin{equation}\label{uni}
\frac{d^2 u}{dr^2}+
\left[\,\om^2+\,\frac{2\om^2}{r-1}\,+\,\frac{\om^2+1/4}{(r-1)^2}\,\right]u
= 0\,.
\end{equation}
We have omitted here also the terms $-(j+1/2)^2 +s^2$ and
$-(j+1/2)^2 + 1/2$ in the coefficients at $1/(r-1)$ in (\ref{012})
and (\ref{12}), respectively. Though these terms could be easily
included into the solutions, they would result in corrections to
Im$\,\om_n$ on the order of $1/n$ only, which are negligible as
compared to the leading term $\sim n$.

We retain however the term $1/4$ in the coefficient at $1/(r-1)^2$
in (\ref{uni}). Otherwise the wave function asymptotic for $z \to
-\infty$ would be $e^{-i\omega z+1/2}$, instead of $e^{-i\omega
z}$. In other words, the effective potential in the initial Regge
-- Wheeler equations (\ref{rw0}), (\ref{rw012}) would not vanish
for $z \to -\infty$, but would tend instead to $1/4$. Indeed, the
wave function asymptotic for $z \to -\infty$ is determined by the
discussed coefficient at $1/(r-1)^2$. Since the coefficient
$\om^2+1/4$ at $1/(r-1)^2$ in (\ref{uni}) corresponds to $\om^2$
in equations (\ref{rw0}), (\ref{rw012}), then obviously the
coefficient $\om^2$ in (\ref{uni}) would correspond to $\om^2-1/4$
in (\ref{rw0}), (\ref{rw012}).

To summarize, it is only natural that equation (\ref{uni}),
essentially semiclassical one (due to the assumption $|\,\om| \gg
1)$, is universal, i.e. independent of spin $s$. Moreover, even if
one assumes that $j \gg 1$ as well (i.e. gives up the condition $j
\ll |\,\om|$ used in (\ref{uni})), the resulting, again
semiclassical equation
\begin{equation}\label{uni1}
\frac{d^2 u}{dr^2}+\left\{\om^2 + \,\frac{2\om^2}{r-1}\,  +
\frac{1}{(r-1)^2}\left(\om^2 + \,\frac{1}{4}\right)-
\,\frac{1}{r(r-1)}\left(j+\frac{1}{2}\right)^2 \right\}u=0\,
\end{equation}
is still universal, i.e. spin-independent.

We address now the eigenvalues of equation (\ref{uni}). Its two
independent solutions can be conveniently expressed via the
Whittaker functions $W_{\lambda,\mu}(x)$ (see, e.g., \cite{gr}).
They are
\[
W_{i\omega,i\omega}(-2\,i\,\omega(r-1))\,, \quad
W_{-i\omega,i\omega}(2\,i\,\omega(r-1))\,.
\]
With their different asymptotic for $r \to \infty$,
\[
W_{i\omega,i\omega}(-2\,i\,\omega(r-1)) \to e^{i\omega [r + \ln
(r-1)]} = e^{i\omega z},
\]
\[
W_{-i\omega,i\omega}(2\,i\,\omega(r-1)) \to e^{-i\omega [r + \ln
(r-1)]} = e^{-i\omega z},
\]
these solutions are obviously independent. On the other hand, the
second one does not comply with boundary condition (\ref{bc}) and
therefore should be excluded.

As to the first solution, its limit for $r \to 1$ is
\[
W_{i\omega,i\omega}(-2\,i\,\omega(r-1)) \;\longrightarrow
\]
\beq\label{hor} \longrightarrow \frac{\Ga(-2\,i\,\om)}{\Ga(1/2 -
2\,i\,\om)}\,[- 2\,i\,\om(r-1)]^{i\omega + 1/2} +
\frac{\Ga(2\,i\,\om)}{\Ga(1/2)}\,[- 2\,i\,\om(r-1)]^{-i\omega +
1/2}. \eeq When going over to the function $\Psi$ used in the
``tortoise'' coordinate $\,z$ (see (\ref{tr})), the overall factor
$(r-1)^{1/2}$ in this expression cancels, and $(r-1)^{\pm
i\omega}$ goes over into $e^{\pm i\omega z}$  for $r \to 1$. To
comply with the boundary condition at the horizon, one should get
rid of the first term in equation (\ref{hor}). To this end,
recalling that $\Ga(- n)$ has poles for integer positive $n$, we
put $1/2 - 2\,i\,\om = - n$, or $\om_n =  - \,(i/2)\,(n+1/2)$.

In fact, equation (\ref{uni}) by itself was obtained from
(\ref{012}) and (\ref{12}) under the assumption $|\om_n| \to
\infty$, or $n \gg 1$. Therefore, in this way we can guarantee,
for the initial problem, only that
\beq\label{app}
\om_n = \,-\,\frac{i}{2}\,n\,, \quad n \gg 1\,,
\eeq
for all spins.

Though less accurate than quantization rules (\ref{o02}) and
(\ref{o1}), this one is still quite sufficient for insisting that
the correct leading term in the quantization rule for spin
\,$1/2$\, is (\ref{o12}), but not (\ref{kon}).

In conclusion of this section, we demonstrate, with a relatively
simple example of truncated equation (\ref{uni}), an analytical
approach, that will allow later, for more accurate treatment of
the wave equations, to find not only the leading term $\sim n$ in
the asymptotic of QNMs, but as well the next, constant one. The
method goes back to \cite{mon} where it was applied to the Regge
-- Wheeler equation for $s = 0,\, 2$. After finding in the present
section by this method the eigenvalues of equation (\ref{uni}), we
will apply below the technique to the Teukolsky equation for
arbitrary spins. Our line of reasoning differs from that
of~\cite{mon}.

Equation (\ref{uni}) has two singular points, $r=1$ and
$r=\infty$. We connect them by a cut in the complex plane $r$
going, for instance, from $r=1$ along the real axis to the right
(solid line in Fig.~1).
\begin{figure}
\centering
\includegraphics{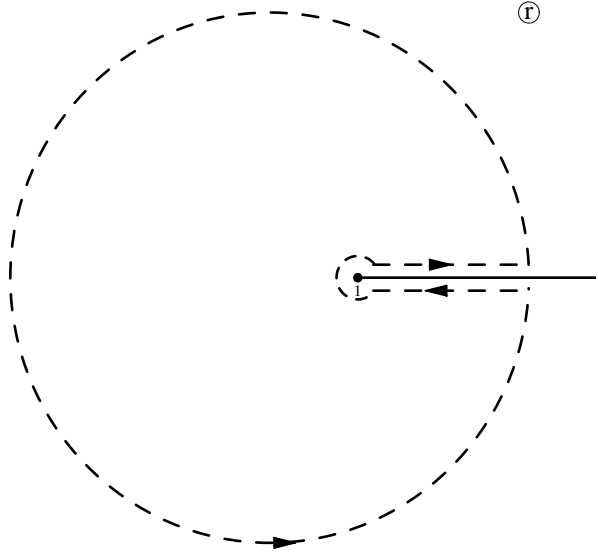}
\caption{Singular point of equation (\ref{uni}), cut, and closed
contour}\label{fig1}
\end{figure}
Let us consider the closed contour marked by the dashed line in
Fig.~1. Since there is no singularity inside it, the solution at
some point on this contour, after going around the contour, comes
back to its initial value, which means that the phase of this
solution changes by $2\,\pi\,n,\; n=0, \pm 1, \pm 2, ....\,$.

When we follow an arc of a large radius $r \gg 1$, where the
asymptotic solution is $e^{i\omega r} r^{i\,\omega}$, i.e. go
around the singular point at infinity, the wave function acquires
the phase $\de(\infty)= 2\pi i \,\omega$.

Then we go around the branch point $r=1$ by following an arc
of a small radius. Here, due to the asymptotic solution
$v(r)=(r-1)^{- i\omega + 1/2}$, the wave function acquires the
phase $\de(1)=2\pi (i\,\omega - 1/2)$. As to the paths along the
cut, they generate no phase at all. Indeed, since $r=1$ is a
regular singular point, the wave function can be written as $u(r)=
v(r)w(r)$, where $w(r)$ is analytic at $r=1$. The phase of
$v(r)=(r-1)^{- i\,\omega + 1/2}$ remains constant along the paths
adjacent to the cut, as well as the phase of $r-1$. As to the
analytic function $w(r)$, it obviously cannot acquire any phase
after going around the cut. In other words, effectively for our
purpose, the regular singular point $r=1$ behaves as if it were an
isolated singularity.

Thus, going counter-clockwise around the considered closed contour
in the complex plane, one obtains
\[
\de(\infty) + \de(1) = 4 \pi i \om - \pi = 2 \pi n\,,
\]
or the quantization rule
\[
\om_n =  - \,\frac{i}{2}\,(n+1/2)\,.
\]
Being interested in the solutions decreasing in time, we choose
here positive $n$ (and of course large ones). Again, one can
guarantee here the leading term only, $\om_n =  - \,i\,n/2$.

\section{Teukolsky equation. Quasinormal modes\\ of odd integer spin}

Now we address the problem of the next, subleading correction, of
zeroth order in $n$, to formula (\ref{app}). It is only natural to
expect that this correction is spin-dependent. So, to investigate
it we will use the Teukolsky equation. As distinct from the Regge
-- Wheeler equation, this one describes in a unified way both
integer and half-integer spins, ranging at least from $s = 0$ to
$s=2$~\cite{teu}--\cite{guv}. Previously, the Teukolsky equation
was used in \cite{lea} for numerical calculations of QNMs.

In the Schwarzschild background the Teukolsky equation for a
massless field is
\begin{equation}\label{teu}
 \Delta \,\frac{d^2 R}{dr^2} +(1-s)(2r-1)\,\frac{dR}{d r}+U(r)R =0\,,
\end{equation}
where
\[
\Delta(r)=r(r-1)\,,\quad U(r)=\,\frac{-\,r(2r-3)\,i\,\omega\,s
+r^3 \omega^2}{r-1}-A_{js}\,,\quad A_{js}=(j+s)(j-s+1)\,.
\]
Obviously, for a given spin $s$ the QNMs are independent of
helicity.

With the tortoise coordinate $z(r)=r+\ln(r-1)$ and new function
$\chi(r)=r \Delta^{-s/2} R(r)$, one obtains the following standard
form for this equation:
\begin{equation}\label{test}
  \frac{d^2\chi}{d z^2}\,+ [\,\om^2 - V(r)]\chi=0\,,
\end{equation}
with the effective potential
\begin{equation}\label{Veff}
  V(r)=\,\frac{s^2-4}{4\,r^4}\, - \,
  \frac{A_{js}^2 - s + s^2-1}{r^3} \,
   + \,\frac{A_{js}^2 - s + s^2 - 3 \,i \,\omega s}{r^2}+ \,
  \frac{ 2\,i \,\omega s}{r}\,.
\end{equation}

Clearly, for $s=0$ $V(r)$ is real, and equation (\ref{test})
coincides with the scalar version of the Regge -- Wheeler equation
(\ref{rw0}). On the other hand, the Teukolsky equation for $s =
1/2$ coincides with the second order equation for a massless Dirac
field in the chiral representation; of course, the latter differs
from equation (\ref{rw012}) written in the standard
representation.

The asymptotic behavior of QNMs of the Teukolsky equation is
\begin{equation}\label{bcte}
\chi \sim \Delta^{s/2}e^{\pm i\,\omega z} \quad {\rm for} \quad
\om z \rightarrow \pm \infty.
\end{equation}

In principle, here the idea of calculating the eigenmodes will be
the same as the second one above, in the case of truncated
equation (\ref{uni}). We choose a closed contour without any
singularity inside it (dashed line in Fig. 2)), calculate the
phase of the wave function acquired after going around the
contour, and equate this phase to $2\pi n,\; n=0, \pm 1, \pm 2,
...\,$. However, this problem for the Teukolsky equation is in
general much more involved than that for the truncated
equation~(\ref{uni})
\begin{figure}
\centering
\includegraphics{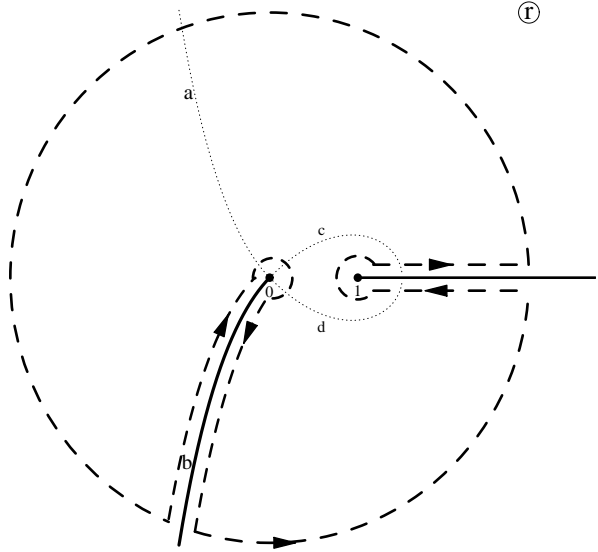}
\caption{Singular points of equation (\ref{test}), cuts, and
closed contour}\label{fig2}
\end{figure}
since equation (\ref{test}) has three singular points: $r=0$,
$r=1$, and $r=\infty$. As previously, we choose a cut in the
complex plane $r$ going from $r=1$ to $r=\infty$ along the real
axis to the right (solid line in Fig. 2). As to the cut starting
at $r=0$ (another solid line in Fig. 2), the choice of its
location will be discussed later. \footnote{Dotted lines in
Fig. 2 are the level lines, see below.}

The treatment of the first cut is practically the same as in the
simple case of equation (\ref{uni}). The present boundary
condition (\ref{bcte}) means that the asymptotic solution for $r
\to 1$ looks here as $(r-1)^{- i\omega + s/2}$. Correspondingly,
when going around the branch point $r=1$ along an arc of a small
radius, the wave function acquires the phase
\beq\label{de1}
\de(1)=2\pi (i\,\omega - s/2)\,.
\eeq
In this case as well, the regular singular point $r=1$ is
effectively equivalent, for our purpose, to an isolated
singularity.

Situation with the cut starting at $r=0$ is more complicated.
The problem is that we have no {\em a priory} boundary condition
at $r=0$, but still have to find the phase acquired when going
around this cut.

However, for $s = 1$ (and for any other odd integer spin that is
described by the Teukolsky equation) the discussed correction can
be found easily. For $r \ll 1$, two independent solutions of
equation (\ref{test}) with potential (\ref{Veff}) are
\beq\label{as0}
\chi_+ \sim\, r^{1\,+\,s/2} \quad {\rm and} \quad \chi_- \sim\,
r^{1\,-\,s/2}\,.
\eeq
Since $r=0$ is a regular singular point of the Teukolsky
equation, the exact general solution of this equation can be
presented as follows:
\beq\label{as1} \chi(r) =
r^{1\,-\,s/2}\left[\sum_{k=0}^{\infty}a_k \,r^k +
r^s\sum_{k=0}^{\infty}b_k \,r^k\right]  =
r^{-\,s/2}\left[\sum_{k=0}^{\infty}a_k \,r^{k+1} +
\sum_{k=0}^{\infty}b_k \,r^{k+s+1}\right].
\eeq
With an odd integer $s$, the singularity of this solution at $r=0$
is due to the overall factor $r^{-\,s/2}$, or to $r^{s/2}$ if by
some reasons $\chi_-$ vanishes (we will see in the next section
that just this is the case). Correspondingly, the phase acquired
by solution (\ref{as0}) as a result of going around the branch
point $r=0$ is
\beq\label{d00}
\de(0) = \pm \pi s\,.
\eeq
In fact, the sign in this expression does not matter for our
problem since the two options differ by $2\pi s$ with an integer
$s$. And if necessary, one can always shift the initial $n \gg 1$
in the quantization rule by an integer $s$.

At last, the asymptotic solution at infinity
\begin{equation}\label{asi}
\chi \sim r^{s + i\omega}e^{i\,\omega \,r}\,, \quad r\rightarrow
\infty\,,
\end{equation}
after going by 2$\pi$ around the arc of infinitely large radius,
acquires the phase
\beq\label{dei} \de(\infty) = 2\pi (i\om +
s)\,.
\eeq
With the phase (\ref{de1}) generated by the branch point $r=1$,
the total result of going around the closed contour is
\[
\de(1) + \de(\infty) + \de(0) = 4 \pi i \om \,
\]
(we have chosen here the sign minus in (\ref{d00})). Equating this
expression, as above, with $2\pi n$, we arrive at the quantization
rule already mentioned in Introduction (see (\ref{o1})), but now
for any odd spin:
\beq\label{qu1}
\om_n = \,-\,\frac{i}{2}\,n \,, \quad s = 1, 3, ...\,.
\eeq

\section{Quasinormal modes of half-integer spin}

The situation is somewhat more complicated for half-integer spins.
Here we have to find out which of the two solutions that behave at
$r \to 0$ as $r^{1+s/2}$ and $r^{1-s/2}$, respectively, is the
true one. To this end we need to match the solution for $|\,r| \ll
1$ to that for $|\,r| \gg 1$. Fortunately, in the limit $|\,\om|
\gg 1$ it can be done analytically (here we follow the idea used
in \cite{mon} for finding the QNMs of Regge -- Wheeler
equation~(\ref{rw0})).

Here and below, to investigate the singularity at $r=0$, it is
convenient to shift $z \to z + i \pi$, so that now
$z(r) = r + \ln(1-r)$, and in the limit $r \ll 1$ we have $z(r) =
-\,r^2/2$. Introducing new variable $\rho=\omega z(r)= -
\,\om\, r^2/2$, we transform equation (\ref{Veff}) in the limit
$|\,\om| \gg 1$ to \footnote{If one retains one more term in the
expansion of $z(r)$, i.e. with $z(r) = -\,r^2/2-\,r^2/3\,$, a
correction
\[
-\,\frac{1-s^2+3j(j+1)}{6\,{\sqrt{2}}\,\rho^{3/2}\omega^{1/2}}\,,
\]
of first order in $|\om|^{-1/2}$, arises in square brackets of
equation (\ref{mq}). It coincides with the corresponding
perturbation obtained in \cite{MAS} that generates corrections
$\sim n^{-1/2}$ to $\omega_n$ for $s=0,2$ in the Regge -- Wheeler
formalism.}
\begin{equation}\label{mq}
  \frac{d\chi^2}{d \rho^2}+\left[1
  -\,\frac{3is}{2 \rho}+\frac{4 - s^2}{16 \rho^2}\right]\chi=0\,.
\end{equation}
Independent solutions of equation (\ref{mq}) are the Whittaker
functions
\beq\label{wf}
W_{\frac{3s}{4},\frac{s}{4}}(-2i\rho) \quad {\rm and} \quad
W_{-\frac{3s}{4},\frac{s}{4}}(2i\rho)\,.
\eeq
Though derived for $|\,r| \ll 1$, these solutions are valid also
for $|\rho|\,=\,|\,\om\, r^2/2\,| \gg 1$, if $|\,\om|$ is
sufficiently large. Their asymptotic behavior for $|\rho|= |\om z|
\to \infty$ is, respectively,
\beq\label{wfa}
\rho^{\frac{3s}{4}} e^{i \rho} \quad {\rm and} \quad
\rho^{-\frac{3s}{4}} e^{-i \rho}\,.
\eeq

To choose the appropriate solution we compare the asymptotic
behavior (\ref{wfa}) with that of the solution of exact equation
(\ref{test}) for $|\om z| \to \infty$, as given in (\ref{bcte}).
With the leading asymptotic for QNMs already established in
section~2 for arbitrary spins, $\om_n \simeq - \,i\,n/2$, the
solution (\ref{bcte}) is exponentially small in the left
half-plane $r$. Therefore, we have to choose here the
exponentially small solution of equation~(\ref{mq}). In this way
we arrive at
\beq\label{wf1}
\chi(\rho)=W_{\frac{3s}{4},\frac{s}{4}}(-2i\rho)\,\sim
\rho^{\frac{3s}{4}} e^{i \rho}.
\eeq

We note that matching of the two solutions, (\ref{bcte}) and
(\ref{wf1}), is not precluded by the fact that the pre-exponential
factors in their asymptotic, $\De^{\frac{s}{2}} \sim (\om z)^s$
and $\rho^{\frac{3s}{4}} \sim (\om z)^{\frac{3s}{4}}$,
respectively, are different. This difference is only natural since
the factor $(\om z)^s$ is due to the term $2\,i \,\omega s/r$ in
equation (\ref{Veff}), and the factor $(\om z)^{\frac{3s}{4}}$ is
due to the term $-3is/2 \rho = 3is/\om r^2$ in equation
(\ref{mq}). The coincidence of the exponentials themselves is
quite sufficient reason to believe that it is just (\ref{wf1})
that reproduces the behavior of the exact solution for $|r| \to
0$.

The functions $W_{\lambda,\,\mu}(y)$ with a given asymptotic
behavior for $|y| \to \infty$ can be expressed via other linearly
independent solutions $M_{\lambda,\,\mu}$ of the Whittaker
equation with a definite behavior for $|y| \to 0$. These solutions
are \cite{gr}
\beq\label{MPh}
M_{\lambda,\,\mu}(y)=y^{\mu+\frac{1}{2}}\,e^{-\frac{y}{2}}\,
\Phi(1/2+\mu-\lambda,1+2\mu,y)\,,
\eeq
where $\Phi(a,b,y)$ is the confluent hypergeometric function.
Functions $W_{\lambda,\,\mu}$ and $M_{\lambda,\,\mu}$ are related
as follows \cite{gr}:
\begin{equation}\label{rel1}
W_{\lambda,\,\mu}(y)=\,\frac{\Gamma(-2\mu)}{\Gamma(1/2-\mu-\lambda)}\,
M_{\lambda,\,\mu}(y)+\,\frac{\Gamma(2\mu)}{\Gamma(1/2+\mu-\lambda)}\,
M_{\lambda,\,-\mu}(y)\,.
\end{equation}
In the present case we have
\begin{equation}\label{rels}
W_{\frac{3s}{4},\frac{s}{4}}(-2i\rho)=\,\frac{\Gamma(-s/2)}{\Gamma(1/2-s)}\,
M_{\frac{3s}{4},\frac{s}{4}}(-2i\rho)+\,\frac{\Gamma(s/2)}{\Gamma(1/2-s/2)}\,
M_{\frac{3s}{4},-\frac{s}{4}}(-2i\rho)\,.
\end{equation}

For half-integer $s$, the first term in this expression vanishes
since $\Ga(z)$ turns to infinity for negative integer $z$. So,
here our solution (\ref{wf1}), with the account for (\ref{MPh})
and with $\rho \sim r^2$, behaves for $r \to 0$ as
\beq
\chi \sim r^{1-s/2}\,.
\eeq
Here again (see (\ref{d00})) the phase due to the branch point
$r=0$ is
\beq\label{d012}
\de(0) = \pi s\,,
\eeq
but now $s$ is half-integer. We have again the same $\de(1)$ and
$\de(\infty)$ as those for odd spin (see (\ref{de1}) and
(\ref{dei}), respectively), and the same quantization condition
\[
\de(1) + \de(\infty) + \de(0) = 2\pi n\,.
\]
At last, shifting the initial $n$ by  the integer part of (now
half-integer) $s$, we arrive at the quantization rule for any
half-integer spin:
\beq\label{qu12}
\om_n = \,-\,\frac{i}{2}\,\left(n + \,\frac{1}{2}\,\right), \quad
s = 1/2 \,, \; 3/2 \,,\; ...\,.
\eeq

Of course, equation (\ref{rels}) can be directly employed also for
odd integer $s$. In this case the second term in this equation
vanishes, and now obvious line of reasoning results in
formula~(\ref{qu1}).

\section{Quasinormal modes of arbitrary spin}

The problem for arbitrary $s$, including the case of direct
physical interest, that of $s = 0,\,2$, requires more
sophisticated approach. In this general case both terms in the rhs
of equation (\ref{rels}) survive, so that the solution near the
origin contains both powers of $r$:
\beq\label{gen}
\chi = a r^{1\,+\,s/2} + b r^{1\,-\,s/2}\,.
\eeq
Obviously, the rotation of this expression by $2 \pi$ around the
branch point $r=0$ in no way can result in its multiplication by
some factor, i.e. the limit (\ref{gen}) of the solution for small
$r$ cannot transform into itself under this procedure.\footnote{An
additional problem arises for even $s$. In this case
$\Gamma(-s/2)$ in equation (\ref{rels}) turns to infinity (and for
$s=0$, $\Gamma(s/2)$ turns to infinity as well), so that this
solution stays finite due to a delicate cancellation between two
terms in the rhs of equation (\ref{rels}). Therefore, to obtain
the explicit form of the solution for even $s$ one should perform
a careful limiting transition. Anyway, this solution for small $r$
also does not transform into itself under the rotation around the
origin.} However, such a transformation does exist for the
solution far away from the origin. The rotation gets possible due
to the Stokes phenomenon, rather well-known in mathematical
physics.

We will not discuss this phenomenon in general, but instead will
demonstrate directly how it works, by solving our problem. We
consider the approximate solution
$W_{\frac{3s}{4},\frac{s}{4}}(-2i\rho)$ for generic $s$ and $\rho$
(though confine of course to small $r$). The analytic continuation
to the specific values of spin will be performed only in the final
result which is a smooth function of $s$.

At first we discuss the position of those lines in the complex
plane $r$ where ${\rm Im}\,\rho = {\rm Im}\,(\omega z) = 0$. These
four level lines in the complex $r$ plane are presented in Fig. 2
(dotted lines therein). Their behavior corresponds to the leading
asymptotic for QNMs, already established in section~2 for
arbitrary spins: $\om_n \simeq - \,i\,n/2$. For small $|\,r|$
(still, $|\rho|\,=\,|\,\om\, r^2/2\,|$ can be large!) the
definition of level lines in the complex $r$ plane reduces to
${\rm Im}(\omega r^2) = 0$, or ${\rm arg}\, r = -\, 1/2\,{\rm
arg}\,\om$. Two of them, $c$ and $d$, going to the right from the
origin, with ${\rm arg}\, r  = \pm \,\pi/4$, are of no special
interest to us. We will be interested mainly in the two level
lines, $a$ and $b$, that become vertical at infinity. In the
sector between these lines the exact solution is exponentially
small for $r\rightarrow \infty$ (see (\ref{asi})).

Our solution (\ref{wf1}) has a cut going from the branch point
$r=0$ to infinity. This cut should be chosen in a judicious,
self-consistent way.

For instance, it cannot be drawn in the sector where the solution
is exponentially small. Indeed, as it was demonstrated above, if
we went around such a cut starting from the small solution
(\ref{wf1}), we would arrive in the result at a linear combination
of both small and large solutions. But a large solution should not
exist in this sector.

By the same reason, if starting from the real positive $r$ axis we
go in the positive direction, i.e. counter-clockwise, along the
contour of large $r$, the cut cannot be drawn along the level line
$a$ or in the sector to the right of it.

Neither, with this direction, should we draw the cut in the sector
between the level lines $b$ and $c$ where the solution is
exponentially large at infinity. In this sector we cannot
guarantee the absence of an exponentially small admixture to the
right of the level line $b$; then, after going around the cut, the
correct solution would be completely distorted.

Thus, for the counter-clockwise direction, the only consistent
choice for the cut is that along the level line $b$. Just in this
way we will proceed. \footnote{Quite analogous arguments
demonstrate that if one goes from the real $r$ axis in the
negative, clockwise direction, the cut should be chosen along the
level line $a$.}

So, let us go from the real $r$ axis in the counter-clockwise
direction along a contour of large $r$. We reach the level line
$b$, and then proceed along its upper side. At a small distance
from the origin, we follow an arc of a radius $r \ll 1$. Then we
come back along the lower side of the cut to the arc of a large
radius $r \gg 1$. At last, we close the contour by going along
this arc, and then around the cut starting at $r=1$.

As to the cut starting from the origin, here again the
contributions to the acquired phase from the upper and lower sides
of the cut cancel. So, we have to find only the phase generated by
the rotation around the branch point $r=0$. The rotation angle
here is $-2\pi$ in the $r$ plane, which corresponds to $-4\pi$ in
the $\rho$ plane. To calculate the mentioned phase, we use
solution (\ref{wf1}) in the limit $\rho \gg 1$, i.e. we work in
the interval $|\,\om\,|^{-1/2} \ll r \ll 1$ (recall that
$|\,\om\,| \gg 1$).

To perform the rotation by $2 \pi$ in the $r$-plane, or by $4 \pi$
in the $\rho$-plane, we note first of all that functions
$M_{\lambda,\,\mu}$ transform under the rotations in a simple way.
Indeed, according to equation (\ref{MPh}),
\[
M_{\lambda,\,\mu}(y\,e^{-4i\pi})=e^{-4i\pi\mu}\,M_{\lambda,\,\mu}(y)\,.
\]

Then we need the relation inverse to (\ref{rel1}), to express back
$M_{\lambda,\,\mu}$ via $W_{\lambda,\,\mu}$. Its form depends on
${\rm arg}\,y$ (see \cite{gr}, 9.233.1, 9.233.2). In our case, the
required initial value of $y = -2i\rho$ corresponds to the upper
side of the cut along the level line $b$. Since ${\rm arg}\,y$
remains constant along a level line, it can be found most easily
for $|r| \to \infty$ where on this line ${\rm arg}\,r = 3\pi/2$.
In such a way, we have here
\[
{\rm arg}\,y = {\rm arg}\,(-2i\rho) =  -\,\frac{\pi}{2}\,+\,{\rm
arg}\,\om\,+ {\rm arg}\,\,r  = \,\frac{\pi}{2}\,.
\]
With this ${\rm arg}\,y$, the inverse relation reads
\[
M_{\lambda,\,\mu}(y)=\,\frac{\Gamma(1+2\mu)}{\Gamma(1/2+\mu-\lambda)}\,e^{
-i\pi \lambda}\,W_{-\lambda,\,\mu}(e^{-i\pi}y) \quad
\quad\quad\quad\quad\quad\quad\quad\quad\quad\quad\quad\quad\quad
\]
\beq\label{inv}
+\,\frac{\Gamma(1+2\mu)}{\Gamma(1/2+\mu+\lambda)}\,e^{i\pi(1/2+\mu-
\lambda)}\,W_{\lambda,\,\mu}(y)\,; \quad -\frac{\pi}{2} < {\rm
arg}\,y < \frac{3\pi}{2}\,.
\eeq

Now we go back in the rotated solution
\[
\chi(\rho\, e^{-4i\pi})
=\,\frac{\Gamma(-2\mu)}{\Gamma(1/2-\mu-\lambda)}
e^{-4i\pi\mu}\,M_{\lambda,\,\mu}(-2i\rho)\,\quad\quad\quad\quad
\]
\beq
+\,\frac{\Gamma(2\mu)}{\Gamma(1/2+\mu-\lambda)}\,
e^{4i\pi\mu}\,M_{\lambda,\,-\mu}(-2i\rho)\,
\end{equation}
to functions $W_{\pm\lambda,\,\pm\mu}(\pm 2i\rho)$ by means of
(\ref{inv}). To simplify the result of this transformation, we
note first of all that the contributions to the result originating
from the first term in equation (\ref{inv}) (containing
$W_{-\lambda,\,\mu}(e^{-i\pi}y)$), are exponentially small along
all the path leading from the line $b$ to the real axis; besides
they vanish of course on this axis. So, these terms can be
neglected at all.\footnote{In the region where $|y| \gg 1$ and the
asymptotic form of the function is used, we in fact have neglected
already small power-like corrections to this form. So much the
more, we can and even should neglect exponentially small
corrections to it. This is the Stokes phenomenon at work.} Then,
we need the result only in the limit $|\rho| \gg 1$, where
$W_{\lambda,\,\mu}(y)=W_{\lambda,\,-\mu}(y)$. In this way, we
arrive at relation
\beq\label{r0}
\chi(\rho\, e^{-4i\pi})= -\,e^{-i \pi s}(1+2\cos\pi s)
\,W_{\frac{3s}{4},\frac{s}{4}}(-2i\rho)\,.
\eeq
The coefficient $\;-\,e^{-i \pi s}(1+2\cos\pi s)\,$ here results
from trivial, but rather tedious transfor\-ma\-tions with
$\Ga$-functions, sines, and cosines. This coefficient can be
rewritten as $\,e^{i\,\delta(0)}\,,$ where
\begin{equation}\label{de0}
\delta(0) = \pi  -\pi s -i \ln(1+2\cos\pi s)
\end{equation}
is the phase acquired by the solution when following the arc of a
small radius $r \ll 1$ around the origin $r = 0$.

As usual, the quantization condition for $\omega_n$ is
\begin{equation}\label{W}
\de(1) + \de(\infty) + \de(0)=2\pi n \,.
\end{equation}
Finally, with (\ref{de1}), (\ref{dei}), and (\ref{de0}), we obtain
analytically the universal formula
\beq \label{fin}
\om_n = \, -\, \frac{i}{2}\left(n+\,\frac{1}{2}\right)\,
+\,\frac{1}{4\pi}\,\ln(1+2\cos\pi s)\,, \quad n \to \infty
\eeq
for eigenmodes of any spin $s$ described by the Teukolsky
equation.\footnote{Of course, the same result arises when going in
the opposite direction along the contour in the complex $r$ plane,
with a cut made along the level line $a$. However, in this case
equation (\ref{inv}) modifies to:
\[
M_{\lambda,\,\mu}(y)=\,\frac{\Gamma(1+2\mu)}{\Gamma(1/2+\mu-\lambda)}\,e^{
i\pi \lambda}\,W_{-\lambda,\,\mu}(e^{i\pi}y)
+\,\frac{\Gamma(1+2\mu)}{\Gamma(1/2+\mu+\lambda)}\,e^{-i\pi(1/2+\mu-
\lambda)}\,W_{\lambda,\,\mu}(y)\,;\] \[ -\frac{3\pi}{2} < {\rm
arg}\,y < \frac{\pi}{2}\,.\]}

\section{Conclusions}

Now few comments on our results.

It is not clear whether the Teukolsky equation is valid for $s >
2\,$.

For $s=0,2$ formula (\ref{fin}) was derived previously in
\cite{mot,mon} in the Regge -- Wheeler formalism. For these spins
it gives
\[
\om_n = \, -\, \frac{i}{2}\left(n+\,\frac{1}{2}\right)\,
+\,\frac{1}{4\pi}\,\ln 3\,, \quad n \to \infty\,.
\]

The result
\[
\om_{\it n} = \, -\, \frac{i}{2}\, n\,,  \quad n \to \infty\,,
\]
for $s=1$ was previously obtained in \cite{mot,car}. It is derived
in an elementary way in section~3, confirmed in section 4, and
follows immediately from (\ref{fin}).

For half-integer spins, simple calculation in section 4, as well as
formula (\ref{fin}), give
\[
\om_{\it n} = \, -\, \frac{i}{2}\left(n+\,\frac{1}{2}\right),
\quad n \to \infty\,, \quad s\,=\,1/2\,,\,3/2\,.
\]
For $s=1/2$ it not only confirms the conclusion of \cite{jin},
thus resolving the controversy on the matter, but contains also
first nonvanishing correction to the leading term; the result for
this correction is new. Our total result for $s=3/2\,$ is also new.
Quite recently, both results were confirmed in \cite{cho}.

\section*{Acknowledgements}
We are grateful to V.M. Khatsymovsky, A.A. Pomeransky, and V.V.
Sokolov for their interest to this work and useful discussions.
The investigation was supported in part by the Russian Foundation
for Basic Research through Grant No. 05-02-16627.

\end{document}